\documentstyle[preprint,eqsecnum,aps]{revtex}
\begin{document}
\draft
\preprint{KYUSHU-HET/021}
\title{
   Dynamics of the Light-Cone Zero Modes:\\
   Theta Vacuum of the Massive Schwinger Model
}
\author{ 
    Koji Harada, 
    Atsushi Okazaki, 
    and Masa-aki Taniguchi
}
\address{
        Department of Physics, Kyushu University\\
        Fukuoka 812-81, Japan
}

\date{March 19, 1996}
\maketitle
\begin{abstract}
The massive Schwinger model is quantized on the light cone with
great care on the bosonic zero modes by putting the system in a 
finite (light-cone) spatial box. 
The zero mode of $A_{-}$ survives Dirac's procedure for the constrained 
system as a dynamical degree of freedom. 
After regularization and quantization, we show that the physical space 
condition is consistently imposed and 
relates the fermion Fock states to the
zero mode of the gauge field.
The vacuum is obtained by solving a Schr\"odinger equation in a 
periodic potential, so that the theta is understood as the Bloch
momentum.
We also construct a one-meson state in the fermion-antifermion sector
and obtained the Schr\"odinger equation for it.
\end{abstract}
\pacs{11.10.Kk,11.10.St,11.15.Tk}

\narrowtext
\section{Introduction}
\label{intro}
In a recent series of papers\cite{hsty,HOT,HOT2}, 
we have examined the massive Schwinger 
model in the Light-Front Tamm-Dancoff (LFTD) 
approximation\cite{phw90} and have 
obtained some interesting non-perturbative results 
which have never obtained by other 
methods, such as bosonization and lattice simulations. The power of
the LFTD approximation suggests that, once it is applied to 
$QCD_{1+3}$, we would have remarkable progress in the study of 
relativistic bound state problems. 

The LFTD approximation is the Tamm-Dancoff approximation\cite{td}
(a truncation of the infinite dimensional Fock space 
by limiting the number of constituents ) applied to field theory 
quantized on the light cone. The light-cone quantization is essential 
for the Tamm-Dancoff approximation: the vacuum is quite 
simple in the light-cone quantization because pair creations 
(annihilations) from (into) the vacuum 
are kinematically suppressed.

How does a complex vacuum structure emerge in light-front field 
theory with the simple vacuum? Actually this is one of the most 
important questions in light-front field theory\cite{bmpp}. 
(In our previous 
papers, we completely neglected the vacuum structure.)
In this paper, we will discuss the simplest non-trivial example of 
the vacuum structure which likely leads to observable effects,
the theta vacuum of the massive Schwinger model. 

The massive Schwinger model has been studied by many 
authors\cite{cjs75,col76}
because it shares several important features with  
$QCD_{1+3}$ such as quark confinement, anomalous $U(1)_A$ breaking 
as well as $\theta$-vacuum. 
In his seminal paper\cite{col76}, 
Coleman showed that the vacuum angle $\theta$ 
can be regarded as an external 
constant electric field. One of his important results is that 
the periodicity of physics in $\theta$ is 
a consequence of dynamical structure of 
the vacuum. Namely, it comes from the 
fact that a pair creation of a fermion and an anti-fermion 
from the vacuum is 
energetically favorable in a background electric field stronger than a 
certain critical value.

 How can this dynamical feature of $\theta$ be understood in the light-cone 
quantization with a simple vacuum? The most important result of the present 
paper is the demonstration that the dynamics of the zero mode 
of the gauge field is responsible for it. 
In order to explicitly extract the zero modes, we first put the system 
into a finite 
light-cone spatial box ($x^-\in [-L,L]$) and impose the periodic boundary 
condition \cite{my76,pb}, 
keeping in mind that we should eventually take the limit 
$L\rightarrow\infty$. (We neglect the fermion zero modes at 
all, by choosing the anti-periodic boundary condition for the 
fermionic variables.)
Even after fixing a gauge, the zero mode of the longitudinal component 
of the gauge field ($\stackrel{o\ }{A_{-}}$) 
remains to be dynamical while the other gauge components 
($\tilde{A}_{-}$,$\stackrel{o\ }{A_{+}}$,$\tilde{A}_{+}$) do not.
We show that we can impose the physical state condition (the 
chargeless condition) consistently at the quantum level. It relates the
fermion Fock states to the zero mode of the gauge field. 
The theory is still invariant under large gauge transformations 
($\pi_{1}(S^{1})=Z$). We look for states which satisfy the 
physical state condition and are invariant under large gauge 
transformations. The vacuum is 
obtained by solving a simple Schr\"odinger equation in a periodic 
potential. The theta variable is identified as the Bloch momentum.
We also obtain the Schr\"odinger equation for the meson state 
in the fermion-antifermion sector.

There are several papers on the massive Schwinger model on the light 
cone. Bergknoff \cite{berg77} first applied the LFTD approximation and Mo
and Perry\cite{mp93}  refined his calculations 
by using the method of basis functions. We have achieved six-body LFTD 
calculations in order to investigate two-meson and three-meson bound 
states \cite{HOT}.
Eller, Pauli and Brodsky \cite{epb87} considered the discretized
light-cone quantization  (DLCQ) of the massive Schwinger model.
The present paper is based on  these. We refer the readers to
them.

The theta term has not been discussed much in the light-cone context. 
Although there are several papers on the theta term (vacua) 
in the {\it massless} Schwinger 
model\cite{McCartor}, the {\it massive} Schwinger model has 
been rarely studied. Our 
approach may appear similar to that of
Heinzl, Krusche and Werner \cite{hkw91}, 
who discussed the zero modes of the gauge field in the massless Schwinger 
model and how the theta vacua arises. There are, however,
critical differences; 
(1) Their treatment of the regularized
current is  not adequate. In their paper the chiral anomaly 
is {\it not} derived through the  point-splitting regularization 
but as a consequence of the {\it classical} equation of motion. 
(2) They impose a regularized charge
density as an  additional constraint, which leads to a {\it second-class}
zero-mode Gauss law (the chargeless condition).
(3) The zero mode 
of the gauge field can take only certain discrete values.  
Of course these are different from
the usual treatment of the regularized currents and the Gauss law. 
In the present paper, 
we regularize the current by point-splitting with a path-dependent phase
factor to make it gauge invariant and examine the structure of constraints.
We find that the regularization does not affect the structure 
of constraints.
We end up with the first-class (zero-mode) Gauss law and 
the usual chiral anomaly which
arises from a gauge-invariant regularization procedure.
The zero mode of the gauge field can take any value without 
inconsistency.

We emphasize that our approach is {\it not} the one in which the gauge 
field is treated as an external field. The (zero mode of the) gauge 
field is treated as a full-fledged quantum-mechanical dynamical degree
of freedom. We think that this distinguishes our work from most of the 
previous papers in which it is never clear whether the (zero mode of 
the) gauge field is an external $c$-number field or not. In order to 
keep our formulation as transparent as possible, we consider the 
states for the {\it whole} system. We can define the conserved, 
gauge-invariant
charge, which in this case 
depends on the quantum-mechanical gauge field degree of freedom.
By using the charge, we succeed in imposing
physical state condition consistently.
The physical state condition plays 
a crucial role in combining the zero mode and the fermion Fock states.

In Sec.~{II}, we first examine the constraints and 
eliminate dependent degrees
of freedom, paying attention to bosonic zero modes. 
We find that the zero
mode of $A_-$ and its canonically conjugate 
momentum remain to be dynamical.
To quantize the theory, we need to regularize operators 
to make them 
well-defined. We show that we can define the conserved charge which 
does not affect the classical structure
of constraints. 
Interestingly, the physical state condition relates the fermion Fock 
states to the zero mode of the gauge field.

In Sec.~{III}, we investigate the physical states. 
We first consider the ground state. By imposing the 
physical state condition
and that it is an eigenstate of the light-cone Hamiltonian, we can 
derive a Schr\"odinger equation (the vacuum equation) in a periodic 
potential, which determines
the vacuum state of the massive Schwinger model in the presence of 
theta. The theta can be identified as the Bloch momentum. The 
periodicity of theta is self-evident. 
We also consider the two-body eigenstate, the meson. The 
Schr\"odinger equation is derived and the so-called ``continuum 
limit'' (more properly, the thermodynamical limit) 
is discussed. 

Sec.~{IV} is devoted to the conclusion and discussions. We give a detailed
discussion on current regularization and anomaly in Appendix.

\section{Light-cone quantization in a box}
\subsection{constraints of the massive Schwinger model on the light cone}
\label{constraints}

In this section we analyze the classical structure of constraints of 
the massive Schwinger model, 
including zero modes, and derive the Hamiltonian in 
a canonical way. In order 
to explicitly separate the zero modes from the non-zero modes 
of the bosonic variables,
we put the system into a finite light-cone spatial box
($x^- \in [-L,L]$) with the periodic boundary condition\cite{my76,pb}.
{}For the fermionic variables we impose the anti-periodic 
boundary condition 
and disregard their zero-modes completely. We will discuss possible
consequences of the inclusion of the fermionic zero modes in Section 
\ref{discussion}.   

The Lagrangian density of the massive Schwinger model
is given by
\begin{eqnarray}
   \cal L&=&-{1\over4}F_{\mu\nu}F^{\mu\nu}
            +\bar{\psi}
            \left[\gamma^{\mu}(i\partial_{\mu}-eA_{\mu})-m\right]\psi \\
         &=&{1\over 2}(\partial_+ \stackrel{o\ }{A_{-}})^2
            +{1\over 2}(\partial_+\tilde{A}_{-}-\partial_-\tilde{A}_+)^2  
         +\sqrt2(\psi_R^\dagger i\partial_+\psi_R 
            +\psi_L^\dagger i\partial_-\psi_L) \nonumber \\
         & &{}-m(\psi_R^{\dagger}\psi_L
              +\psi_L^{\dagger}\psi_R)
            -e(\sqrt2\psi_R^\dagger\psi_R A_{+}
              +\sqrt2\psi_L^\dagger\psi_L A_{-}),  
\end{eqnarray}
where $\stackrel{o\ }{A_{\pm}}$ stands for 
the zero mode of $A_{\pm}$, $\stackrel{o\ }{A_{\pm}}={1\over 
2L}\int^{L}_{-L}dx^{-}A_{\pm}$, and $\tilde{A}_{\pm}$ the non-zero mode,
$\tilde{A}_{\pm}\equiv A_{\pm}- \stackrel{o\ }{A_{\pm}}$. 
We will use similar notations hereafter. For other notations, we refer 
the readers to the previous paper\cite{HOT}. \\
The conjugate momenta are obtained as follows:
\begin{eqnarray}
        \stackrel{o}{\pi}^+ &\equiv& 2L\stackrel{o\ }{E^{+}}\approx 0 , \ \ 
         \tilde{\pi}^+ \equiv \tilde{E}^+\approx 0 , \\
        \stackrel{o}{\pi}^- &\equiv& 2L\stackrel{o\ }{E^{-}}
         =2L\partial_+\stackrel{o\ }{A_{-}} ,\ \  
        \tilde{\pi}^- \equiv \tilde{E}^- =
                  \partial_+\tilde{A}_{-}-\partial_-\tilde{A}_+ \\
        \pi^{\dagger}_R &=&i\sqrt2\psi_R^{\dagger} ,\ \   
        \pi_R\approx 0 , \ \  
        \pi^{\dagger}_L\approx 0 , \ \  
        \pi_L\approx 0 .
\end{eqnarray}
Note that because the zero modes do not depend on $x^-$, it is 
useful to extract the factor $L$ from the conjugate momenta, 
and that, for the fermionic 
variables, the daggered/undaggered momenta are conjugate to 
undaggered/daggered variables 
respectively, e.g., 
$\pi_{R}^{\dagger}\equiv \delta L/\delta(\partial_+\psi_R) $. \\
{}From these we see that the primary constraints\cite{dirac} are as follows:  
\begin{equation}
        \theta_1\equiv\stackrel{o\ }{E^{+}}, \ \ 
        \theta_2\equiv\tilde{E}^+ ,\ \ 
        \theta_3\equiv\pi^{\dagger}_R-i\sqrt2\psi_R^{\dagger} ,\ \  
        \theta_4\equiv\pi_R ,\ \  
        \theta_5\equiv\pi_L^{\dagger} ,\ \  
        \theta_6\equiv\pi_L . 
\end{equation}
The total Hamiltonian becomes 
\begin{eqnarray}
        H&=&\int^{L}_{-L}dx^- 
        [
        {1\over2}(\stackrel{o\ }{E^{-}})^2+{1\over2}(\tilde{E}^-)^2
        +\tilde{E}^-\partial_- \tilde{A}_+
        -\hspace{-.1cm}\sqrt2\psi_L^\dagger i\partial_-\psi_L \nonumber \\
        &+&m(\psi_R^{\dagger}\psi_L+\psi_L^{\dagger}\psi_R)
        +e(\sqrt2\psi_R^\dagger\psi_R A_++\sqrt2\psi_L^\dagger\psi_L A_{-})
        +\sum_{i=1}^6\theta_i\lambda^i
        ] , 
\end{eqnarray}
where $\lambda_i\ (i=1,\dots,6)$ are Lagrange multipliers. 
The consistency conditions for $\theta_3$ and $\theta_4$ only 
determine the 
Lagrange multipliers $\lambda^4$ and $\lambda^3$ respectively. 
The rest lead to  
further (secondary) constraints. 
\begin{eqnarray}
        \varphi_1&\equiv&
                {1\over2L}\int_{-L}^L dx^{-} 
                \sqrt2 e\psi_R^{\dagger}\psi_R(x) ,\\
        \varphi_2&\equiv&
                \partial_-\tilde{E}^{-} 
                -\sqrt2 e(\psi_R^{\dagger}\psi_R(x))_\sim ,\\
        \varphi_5&\equiv&
                i\partial_-\psi_L^{\dagger}+
                {m\over\sqrt2} \psi_R^{\dagger}+e\psi_L^{\dagger} A_{-} ,\\
        \varphi_6&\equiv&i\partial_-\psi_L-
                {m\over\sqrt2} \psi_R-e A_{-}\psi_L .\label{psiconstraint}
\end{eqnarray}
The consistency conditions for these constraints do not lead to any
further constraints. (The consistency conditions of $\varphi_5$ and
$\varphi_6$ determine the multipliers $\lambda_6$ and $\lambda_5$ 
respectively,
while those of $\varphi_1$ and $\varphi_2$ are satisfied automatically.)
As usual we can arrange these constraints 
into first- or second-class ones. 
We find the following first-class constraints,
\begin{eqnarray}
        \theta_1&=&\stackrel{o\ }{E^{+}} ,\ \ 
        \theta_2=\tilde{E}^+ \\
        \varphi_1&=&{-ie\over2L} \int_{-L}^L dx^{-} (\pi_R^{\dagger}\psi_R(x)
        +\psi_R^{\dagger}\pi_R(x)+\pi_L^{\dagger}\psi_L(x)
        +\psi_L^{\dagger}\pi_L(x)), 
        \\
        \varphi_2&=&\partial_-\tilde{E}^- +ie
        (\pi_R^{\dagger}\psi_R(x)
        +\psi_R^{\dagger}\pi_R(x)
        +\pi_L^{\dagger}\psi_L(x)
        +\psi_L^{\dagger}\pi_L(x))_{\sim}  . 
\end{eqnarray}
We choose the following gauge-fixing conditions,
\begin{equation}
        \chi_1\equiv\stackrel{o\ }{A_{+}}\approx0 ,\ \ 
        \chi_2\equiv\tilde{A}_- \approx0,\ \ 
        \chi_3\equiv\tilde{E}^-+\partial_-\tilde{A}_+\approx0 . 
\end{equation}
Note that the consistency of $\chi_2$ gives the third constraint 
$\chi_3$. 
The consistency of $\chi_1$ and $\chi_3$ determine 
the multiplier $\lambda_1$
and $\lambda_2$ respectively. Interestingly one cannot choose 
$\stackrel{o\ }{A_-}\approx0$ because it does not have non-vanishing
Poisson brackets with any of the first-class constraints. We end up
with a single first-class constraint $\varphi_1$, the charge.
We will impose it as a physical state condition after quantization,
\begin{equation}
        \varphi_1|{\rm phys}\rangle=0, 
        \label{psc}
\end{equation}
which eliminates charged states from the physical space.

We use second-class constraints to eliminate dependent degrees of freedom. 
It is easy to see that the independent variables are 
$\stackrel{o\ }{A_{-}}$,\ 
$\stackrel{o\ }{E^{-}}$, \ $\psi_R$ and $\psi_R^{\dagger}$. 
Non-vanishing Dirac brackets\cite{dirac} for these variables are 
calculated as  
\begin{equation}
        \{\psi_R(x^-),\psi_R^{\dagger}(y^-)\}_D =
        {-i\over\sqrt2}\delta(x^--y^-) ,\ \ 
        \{\stackrel{o\ }{A_{-}},\stackrel{o\ }{E^{-}}\}_D ={1\over 2L} . 
\end{equation}    
In terms of independent degrees of freedom, 
the Hamiltonian can be written as
\begin{eqnarray}
        P^-&=&P^-_{zero}+P^-_{fmass}+P^-_{current} ,\label{clham}\\
        P^-_{zero}&=&{L}(\stackrel{o\ }{E^{-}})^2 , \\ 
        P^-_{fmass}&=&{m^2\over \sqrt2}\int^{L}_{-L} dx^-
                   [\psi_R^{\dagger}(x^-)e^{-ie\stackrel{o\ }{A_{-}}x^-} 
                   {1\over i\partial_-} 
                   e^{ie\stackrel{o\ }{A_{-}}x^-}\psi_R(x^-)] ,\\
 P^-_{current}&=&{e^2\over 2}\int^{L}_{-L}dx^- \tilde{j}^+(x^-)
    \left({1\over i\partial_-}\right)^2 \tilde{j}^+(x^-), 
\end{eqnarray}
where the inverse of the derivative operator is understood as the principal
value in the Fourier transforms\cite{zh1}.
Note that the dynamical zero modes ($\stackrel{o\ }{A_{-}}$, 
$\stackrel{o\ }{E^{-}}$) come into the expression in a nontrivial way. 
The first term $P^-_{zero}$ is the energy of the constant electric
field. The second term $P^-_{fmass}$ contains the interaction 
of the zero mode of the gauge field with the fermion, 
and requires a special care. It is interesting to note 
that only the non-zero mode of the current appears 
in the third term $P^-_{current}$.

\subsection{regularization of composite operators, 
charge and subsidiary condition}
\label{quantization}

In order to quantize the theory, we replace Dirac brackets with the 
corresponding ($(-i)$ times) equal-$x^+$ commutators. In addition,
we need to regularize composite operators to make them well-defined.
In two dimensions, one can eliminate all divergences by normal-ordering.
In the following, we carefully define 
the current operators, Hamiltonian, 
and charge in a well-defined way so that the structure
of constraints analyzed in the previous subsection is not altered
by the regularization.

{}First of all, we have to define the ``normal-ordering.''
{}For this purpose,
we treat the gauge field $\stackrel{o\ }{A_-}$ 
(or, $q\equiv (L/\pi)e\!\stackrel{o\ }{A_-}$, which is nothing but the 
Chern-Simons term in one dimension) as an external field for a while 
and quantize the
fermionic variables in this external field.

We Fourier expand the fermionic variable $\psi_R$,
\begin{equation}
        \psi_R(x)={1\over 2^{1/4}\sqrt{2L}}
        \sum_{n=-\infty}^{\infty}a_{n+{1\over2}}
        e^{-i{\pi\over L}(n+{1\over2})x^-}.
        \label{psifourier}
\end{equation}
{}From the corresponding Dirac brackets, $a_{n+{1\over2}}$ is assumed
to satisfy the
following anti-commutation relations, 
$\{a_{n+{1\over2}},a^\dagger_{m+{1\over2}}\}=\delta_{n,m}$, 
and $\{a_{n+{1\over2}},a_{m+{1\over2}}\}=\{a^\dagger_{n+{1\over2}},
a^\dagger_{m+{1\over2}}\}=0$. Using these operators, we define a set of
reference states, so-called ``$N$-vacua,'' in analogy of Dirac 
sea,
\begin{equation}
        |0\rangle_N 
   \equiv \prod_{n=-\infty}^{N-1}a_{n+{1\over2}}^{\dagger}|0\rangle ,
\label{nvac}
\end{equation}
where $|0\rangle$ is the `empty' state, i.e., $a_{n+{1\over2}}|0\rangle=0$
for any $n$. At this moment, $N$ is an arbitrary integer. (The use of 
the ``$N$-vacua'' is rather standard in the Schwinger model in the 
equal-time quantization. See Refs.~\cite{manton86,im90}.)

We regularize the current by point-splitting. We define the current operator
$j^\mu(x)$ in a gauge invariant way,
\begin{eqnarray}
{j}^{\mu} &=& \lim_{\epsilon\rightarrow 0}{1\over 2}
[\bar{\psi}(x+\epsilon)\gamma^{\mu}\psi(x) 
\exp\{-ie\int_x^{x+\epsilon}dx^{\mu}A_{\mu}\} \nonumber \\
& &\quad{}-\psi(x)\bar{\psi}(x-\epsilon)\gamma^{\mu}
 \exp\{+ie\int^x_{x-\epsilon}dx^{\mu}A_{\mu}\}],
 \label{pointsplitting}
\end{eqnarray}
where only $\stackrel{o\ }{A_-}$ and $\tilde{A}_+$ are non-zero. A 
straightforward calculation shows 
\begin{equation}
j^+(x)=\sqrt{2}:\psi^\dagger_R(x)\psi_R(x):_N+{1\over2L}(N-{q}),
\end{equation}
where
\begin{eqnarray}
\sqrt{2}:\psi^\dagger_R(x)\psi_R(x):_N  
& =&{1\over2L}\Big\{(\sum_{n\ge N}\sum_{m\ge N}
+\sum_{n< N}\sum_{m\ge N}+\sum_{n\ge N}\sum_{m< N}) 
a^\dagger_{n+{1\over2}}a_{m+{1\over2}} \nonumber \\
 & &{}-\sum_{n< N}\sum_{m< 
 N}a_{m+{1\over2}}a^\dagger_{n+{1\over2}}\Big\}
e^{i\pi(n-m)x^-/L}.
\label{normalordered}
\end{eqnarray}

In Appendix, we discuss how to obtain the Schwinger term and the 
anomalous
conservation law of the axial vector current.

A problem arises when we treat zero-modes with care. Because of the relation
$j_5^\mu=-\epsilon^{\mu\nu}j_\nu$, the $(+)$-components of these two currents
coincide. Naively, therefore, the charges should be the same. On the other
hand, because the vector current is conserved and the axial-vector 
current is not conserved
anomalously as well as explicitly, one would expect that the vector charge
is conserved while the axial-vector charge is not. This apparent 
contradiction is resolved formally by thinking that the zero modes (the 
charges) have no direct connection with the non-zero modes. Perhaps an 
elaborate work on zero-modes may explain the precise relation between the 
zero modes and non-zero modes of the currents. At this moment, however, we 
take a pragmatic way and simply ``adjust'' the zero mode (charge) so 
that it satisfies desired properties. 
(See Appendix for the axial-vector charge.)

Because the Hamiltonian does not contain the zero modes of the currents, 
it is free from this ambiguity. What we should do is to regularize 
$P^-_{fmass}$, which is essentially the mass term 
$m\int dx^{-}\bar\psi\psi$ written in terms of the independent fields. 
But there is a rather surprising fact; the mass term $\bar\psi\psi$
is {\it not} invariant under charge conjugation on the light-cone. In 
equal-time quantization, in order to prove the charge conjugation
invariance of the mass term, we use the fact that 
$\psi_{R}$ anti-commutes with $\psi_{L}^{\dagger}$. In light-cone 
quantization, on the other hand, they do not anti-commute,
\begin{equation}
        \{\psi_{R}(x^{-}),\psi_{L}^{\dagger}(y^{-})\}={m\over4\pi}
        \sum_{n}{1\over n+{1\over2}-q}
        e^{-i{\pi\over L}(n+{1\over2})(x^{-}-y^{-})}.
        \label{funny}
\end{equation}
Therefore, if we wish to preserve charge conjugation invariance of
$P^{-}$ at the quantum level, we have to define it in an invariant way.
The simplest way is to replace $\bar\psi\psi$ with 
$(\bar\psi\psi-\psi^{T}(\bar\psi)^{T})/2$, where the superscript $T$ 
stands for transpose. By using this definition in the quantum theory, 
we get\cite{martin}
\begin{eqnarray}
        & &{m^2\over2\sqrt2}\int_{-L}^{L}dx^-\left[\psi_R^\dagger 
        e^{-ie\stackrel{o\ }{A_-}x^-}
        {1\over i\partial_-}
        e^{ie\stackrel{o\ }{A_-}x^-}
        \psi_R
        - e^{-ie\stackrel{o\ }{A_-}x^-}
        \left({1\over i\partial_-}
        e^{ie\stackrel{o\ }{A_-}x^-}
        \psi_R\right)\psi_R^\dagger\right]\nonumber \\
        &=&{m^2L\over 2\pi} \left\{
        \sum_{n\ge N}{1\over n+{1\over2}-{q}}
        a_{n+{1\over2}}^\dagger a_{n+{1\over2}}-
        \sum_{n< N}{1\over n+{1\over2}-{q}}
        a_{n+{1\over2}} a_{n+{1\over2}}^\dagger \right\} \nonumber \\
        & &{}+{m^2L\over 4\pi}\left[\sum_{n<N}{1\over n+{1\over2}-{q}}
        -\sum_{n\ge N}{1\over n+{1\over2}-{q}}\right],
\end{eqnarray}
where the last term may be regularized by using $\zeta$-function.
It is rewritten as
\begin{equation}
{m^2L\over 4\pi}\left[\psi({1\over2}+{q}-N)+\psi({1\over2}-{q}+N)\right]
\end{equation}
after dropping $q$-independent infinity, where $\psi$ is a digamma 
function.

We are now going to discuss a very interesting symmetry. Even after
fixing a gauge, there is a residual symmetry, called ``large'' gauge
symmetry. The theory is invariant under a large gauge transformation $U$,
\begin{eqnarray} 
U\psi_R(x)U^\dagger&=&e^{i{\pi\over L}x^-}\psi_R(x),\\
U\stackrel{o\ }{A_-}U^\dagger&=&\stackrel{o\ }{A_-}-{1\over e}{\pi\over L}.
\end{eqnarray}
In terms of $a_{n+{1\over2}}$ and $\hat{q}$, we have
\begin{eqnarray}
Ua_{n+{1\over2}}U^\dagger&=&a_{n+{3\over2}},\\
U\hat{q}U^\dagger&=&\hat{q}-1. 
\end{eqnarray}
(In order to avoid possible confusions, 
we have used the notation $\hat{q}$ for 
the operator.)
Note that this transformation does not change the gauge conditions, and
the boundary conditions for $\psi_R$ and $A_-$. This transformation 
generates an additive group $Z$ and decreases $q$ by one.

It is easy to prove the following transformation properties,
\begin{eqnarray}
        U|0\rangle_N & = & |0\rangle_{N+1}\\
        U|q\rangle& =& |q+1\rangle \\
        UP^-U^\dagger & = & P^-  \label{hamUinv}\\
        U(\sqrt2:\psi_R^\dagger(x)\psi_R(x):_N)U^\dagger & =
         & \sqrt2:\psi_R^\dagger(x)\psi_R(x):_{N+1} \nonumber\\
         &=&\sqrt2:\psi_R^\dagger(x)\psi_R(x):_N-{1\over 2L}.
\end{eqnarray}

At this point it is useful to introduce $M(q)$, the integer closest to $q$,
i.e.,
\begin{equation} 
-{1\over2} < q-M(q) < {1\over 2}, 
\end{equation}
which transforms in the following way,
\begin{equation}
        UM(\hat{q})U^\dagger=M(\hat{q}-1)=M(\hat{q})-1.
\end{equation}

Let us define the charge operator. As we have explained, we do not require 
that the charge must be just the (light-cone) spatial 
integral of the current.
In fact, it is easy to show that such a ``naive'' definition of the charge
\begin{equation}
        Q_{\rm naive}\equiv 2L\stackrel{o}{\jmath}^+
        =\sum_{n\ge N}a^\dagger_{n+{1\over2}}a_{n+{1\over2}}-
        \sum_{n< N}a_{n+{1\over2}}a^\dagger_{n+{1\over2}}+N-\hat{q}
\end{equation}
does not commute with the Hamiltonian,
\begin{equation}
        [Q_{\rm naive},P^-]=-{ie\over \pi}L\stackrel{o\ }{E^-},
\end{equation}
though it is invariant under a large gauge transformation.

We define the charge in the following way,
\begin{equation}
        Q=\sum_{n\ge N}a^\dagger_{n+{1\over2}}a_{n+{1\over2}}-
        \sum_{n< N}a_{n+{1\over2}}a^\dagger_{n+{1\over2}}+N-M(\hat{q}).
\end{equation}
Note that it is invariant under a large gauge transformation and commutes
with the Hamiltonian. (The momentum operator $\stackrel{o\ }{E^-}$ generates
an infinitesimal translation of the coordinate ${q}$. 
The integer part $M(q)$ is
invariant under such a translation.)

We can now impose the physical state condition,
\begin{equation}
        Q|{\rm phys}\rangle=0.
\end{equation}
Because the charge is conserved and is invariant under large gauge 
transformations, this definition of physical states is gauge
invariant and is consistent under 
(light-cone) ``time'' evolution. 

\section{Physical states}
\subsection{vacuum state}

Let us consider a generic state for the total system,
\begin{eqnarray}
        |\phi\rangle&=&
        \int_{-\infty}^{\infty}dq|q\rangle\langle q|\phi\rangle \nonumber \\
        &=& \int_{-\infty}^{\infty}dq|q\rangle\sum_{\alpha}\phi_{\alpha}(q)
        |\rm{Fock}(\alpha)\rangle
\end{eqnarray}
where $\phi_{\alpha}(q)$ is the wave function for the zero mode in the 
$q$-representation, with $\alpha$ parameterizing fermion Fock 
states. The ket $|\rm{Fock}(\alpha)\rangle$ is a fermion Fock state.

It is convenient to consider $U$ as the product of two operators,
\begin{equation}
        U=U_{f}\otimes U_{g}
        \label{U}
\end{equation}
where $U_{f}$ acts only on the fermion variable, 
$U_{f}a_{n+{1\over2}}U_{f}^{\dagger}=a_{n+{3\over2}}$, and $U_{g}$ 
only on $\hat{q}$, $U_{g}\hat{q}U_{g}^{\dagger}=\hat{q}-1$.

The transformation property of $|\phi\rangle$ under $U$ is easily 
derived:
\begin{eqnarray}
U|\phi\rangle & = & \int dq U_{g}|q\rangle\sum_{\alpha}\phi_{\alpha}(q)
\left(U_{f}|\rm{Fock}(\alpha)\rangle\right)  \nonumber \\
 & = & \int dq|q+1\rangle\sum_{\alpha}\phi_{\alpha}(q)
\left(U_{f}|\rm{Fock}(\alpha)\rangle\right)  \nonumber \\
 & = & \int dq|q\rangle\sum_{\alpha}\phi_{\alpha}(q-1)
\left(U_{f}|\rm{Fock}(\alpha)\rangle\right).
\end{eqnarray}
Keeping this in mind, one may consider the transformation of the 
$c$-number $q$ under the $U$ transformation, $q\rightarrow q-1$.

Let us now consider the ground state. We first notice that the state
$|0\rangle_{N}$ has a smaller energy than that of any of the states 
of the form $\prod_{\{n_{i}\}}a^{\dagger}_{n_{i}+{1\over2}}
\prod_{\{m_{i}\}}a_{m_{i}+{1\over2}}|0\rangle_{N}$, where $n_{i}\ge N$
and $m_{i}\le N-1$. The problem is 
that the state $|0\rangle_{N}$ is not a physical state nor $U$-invariant.
We therefore consider a linear combination of $N$-vacua and seek for 
the conditions under which it satisfies all the desired properties.
Note that the simplification that the ground state is a linear 
combination of $N$-vacua even in the {\it massive} Schwinger model 
comes from the fact that the Hamiltonian causes no pair creation from 
the state $|0\rangle_{N}$. 
In equal-time quantization, on the other 
hand, this cannot occur.

Consider the state $|\ \rangle$,
\begin{equation}
        |\ \rangle=\int dq |q\rangle 
        \sum_{N=-\infty}^{\infty}\psi_{N}(q)|0\rangle_{N},
        \label{first}
\end{equation}
and require that $Q|\ \rangle=0$, 
\begin{equation}
        Q|\ \rangle=\int dq |q\rangle 
        \sum_{N=-\infty}^{\infty}\psi_{N}(q)(N-M(q))|0\rangle_{N}=0.
\end{equation}
This is satisfied when $\psi_{N}(q)=\varphi_{N}(q)\delta_{N,M(q)}$ 
for all integer $N$. It appears that we may have infinitely many 
different functions 
$\varphi_{N}(q)$ for different values of $N$. But, because of the 
delta, each function $\varphi_{N}(q)$ is defined only in the region 
$N-1/2 < q  < N+1/2$. As a whole, we define a single function for 
the whole $q$ region, $-\infty < q < \infty$. Let us call it 
$\varphi(q)$. The only assumption we make is that it is a continuous 
function of $q$. By using it, the state $|\ \rangle$ can now be 
written as
\begin{eqnarray}
        |\ \rangle & = & \int dq|q\rangle
        \varphi(q)|0\rangle_{M(q)}  \nonumber \\
         & = & \sum_{M=-\infty}^{\infty}\int_{M-{1\over2}}^{M+{1\over2}}dq
         |q\rangle\varphi(q)|0\rangle_{M}.   
\end{eqnarray}

We now require that it is an eigenstate of $P^{-}$, $P^{-}|\ 
\rangle=2L\epsilon|\ \rangle$, where $\epsilon$ is the energy density.
In the unit of $e/\sqrt{\pi}=1$, the eigenvalue equation becomes
\begin{equation}
        \left[-{1\over2}{d^{2}\over dq^{2}}+
        {m^{2}\over2}\{\psi({1\over2}+q-M(q))+\psi({1\over2}-q+M(q))\}\right]
        \varphi(q)
        =\tilde\epsilon\varphi(q),
        \label{vaceq}
\end{equation}
where $\tilde\epsilon\equiv 4\pi\epsilon$.
It is a Schr\"odinger equation in a periodic potential.
We call this Schr\"odinger equation (\ref{vaceq}) the vacuum equation.

It is well-known as Bloch theorem\cite{Kittel} 
that the solution $\varphi(q)$
of a Schr\"odinger equation for a 
periodic potential (with period $1$) can be written 
in the following form,
\begin{equation}
        \varphi_{\theta}(q)=e^{-i\theta q}\phi(q)
        \label{bloch}
\end{equation}
where $\phi(q)$ is a periodic function. The parameter $\theta$ has been 
introduced as a Bloch momentum.
The periodicity of $\theta$ is evident.

It is not difficult to see that the above $\theta$ is the vacuum angle.
Consider the vacuum state,
\begin{eqnarray}
        |\theta\rangle &=&\int^{\infty}_{-{\infty}}dq
        \varphi_{\theta}(q)     |q\rangle|0\rangle_{M(q)} \nonumber \\
        &=& \sum_{M=-\infty}^{\infty}\int_{M-{1\over2}}^{M+{1\over2}}dq
        \varphi_{\theta}(q)|q\rangle|0\rangle_{M}
        \label{thetavac}
\end{eqnarray}
where the wave function $\varphi_{\theta}(q)$ is the eigenfunction of
the vacuum equation (\ref{vaceq}) corresponding to 
the lowest eigenvalue $\tilde\epsilon_{0}(\theta)$. It is the 
eigenstate of $U$,
\begin{eqnarray}
    U|\theta\rangle 
    &=& \sum_{M=-\infty}^{\infty}\int_{M-{1\over2}}^{M+{1\over2}}dq
        \varphi_{\theta}(q)U_{g}|q\rangle U_{f}|0\rangle_{M} \nonumber \\
        &=& \sum_{M=-\infty}^{\infty}\int_{M-{1\over2}}^{M+{1\over2}}dq
        \varphi_{\theta}(q)|q+1\rangle|0\rangle_{M+1} \nonumber \\
        &=& \sum_{M=-\infty}^{\infty}\int_{M-{1\over2}}^{M+{1\over2}}dq
        \varphi_{\theta}(q-1)|q\rangle|0\rangle_{M} \nonumber \\
        &=& e^{i\theta}|\theta\rangle,
\end{eqnarray}
where we have used eq. (\ref{bloch}) in the last step.
The fact that an eigenstate of $P^{-}$ is also an eigenstate of $U$ 
is a direct consequence of (\ref{hamUinv}).

Note that, though it is known that theta {\it is analogous to} a Bloch
momentum\cite{Jackiw}, what we have shown is that theta {\it is} a 
Bloch momentum in the massive Schwinger model, with the explicit 
periodic potential.

In order to satisfy the requirement $P^{-}|\theta\rangle=0$, one 
has to renormalize the energy,
\begin{equation}
        P^{-}\rightarrow 
        P^{-}_{\theta}=P^{-}-{L\over 2\pi}\tilde\epsilon_{0}(\theta).
        \label{e_renorm}
\end{equation}
This is essential to investigate the meson, which we are now going to 
discuss.
 
\subsection{meson state}
\label{mesonsec}
In this section, we proceed to consider the meson state, by 
approximating it as a two-body state. For this purpose, let us first 
discuss the momentum operator $P^{+}$. Naively, the momentum operator 
is defined as
\begin{eqnarray}
        P_{\rm naive}^{+} & = & \sqrt{2}\int 
        dx^{-}\psi_{R}^{\dagger}i\partial_{-}\psi_{R} \nonumber\\
         & = & {\pi\over L}\left\{
         \sum_{n\ge N}(n+{1\over2})a_{n+{1\over2}}^{\dagger}a_{n+{1\over2}}
         -\sum_{n< N}(n+{1\over2})a_{n+{1\over2}}a_{n+{1\over2}}^{\dagger}
         +\sum_{n<N}(n+{1\over2})\right\}  \label{naivemom}\\
         & = & {\pi\over L}\left\{
         \sum_{n\ge N}(n+{1\over2})a_{n+{1\over2}}^{\dagger}a_{n+{1\over2}}
         -\sum_{n< N}(n+{1\over2})a_{n+{1\over2}}a_{n+{1\over2}}^{\dagger}
         +{N^{2}\over2}-{1\over 24}\right\},\nonumber
\end{eqnarray}
where we have used the formula 
$\sum_{n<N}(n+{1\over2})=-\zeta(-1,{1\over2}-N)$. It is interesting 
to note that, although the above expression is normal-ordered with 
respect to $|0\rangle_{N}$, the expression normal-ordered with respect 
to $|0\rangle_{N+1}$ has the same form (with $N$ replaced by $N+1$.)

What we want to define is the momentum operator which satisfies (i) 
$[P^{+},P^{-}_{\theta}]=0$, (ii) $[P^{+},Q]=0$ and (iii) 
$UP^{+}U^{\dagger}=P^{+}$. The naive operator (\ref{naivemom}) 
satisfies the first two requirements, but does not the third;
\begin{equation}
        UP_{\rm naive}^{+}U^{\dagger}=P_{\rm naive}^{+}-{\pi\over 
        L}\left(Q+M(\hat{q})-{1\over2}\right).
\end{equation}
Is it possible to define such a $P^{+}$ by amending
 $P^{+}_{\rm naive}$? 
 It is not difficult to see the momentum operator $P^{+}$
 defined by
 \begin{equation}
        P^{+}={\pi\over L}\left\{
         \sum_{n\ge N}(n+{1\over2})a_{n+{1\over2}}^{\dagger}a_{n+{1\over2}}
         -\sum_{n< N}(n+{1\over2})a_{n+{1\over2}}a_{n+{1\over2}}^{\dagger}
         -{1\over 2}(N-M(\hat{q}))^{2}
         -M(\hat{q})Q\right\}
        \label{mom}
\end{equation}
satisfies the third requirement,
 \begin{equation}
        UP^{+}U^{\dagger}=P^{+},
 \end{equation}
 without failing to satisfy the first two requirements.
We have already renormalized it so that $P^{+}|\theta\rangle=0$.

It is useful to introduce $b_N^{\dagger}(n+{1\over2})$ 
and $d_N^{\dagger}(n+{1\over2})$ as
\begin{eqnarray}
b_N^{\dagger}(n+{1\over2}) & \equiv & a_{N+n+{1\over2}}^{\dagger}\nonumber \\
d_N^{\dagger}(n+{1\over2}) & \equiv & a_{N-n-{1\over2}},
\end{eqnarray}  
where $n$ is a positive integer. They act as creation operators 
of a fermion and an anti-fermion with respect to the $N$-vacuum 
respectively. It is easy to see that they satisfy 
the anti-commutation relation,  
$\{b_N^{\dagger}(n+{1\over2}),b_N(n'+{1\over2})\}
=\{d_N^{\dagger}(n+{1\over2}),d_N(n'+{1\over2})\}=\delta_{n,n'}$.

In terms of these new operators, $\psi_{R}$, $Q$, and $P^{+}$ are 
rewritten as
\begin{eqnarray}
        \psi_{R}(x) & = & {e^{-i{\pi\over L}Nx^{-}}\over2^{1/4}\sqrt{2L}}
        \sum_{n\ge0}\left\{
        b_{N}(n+{1\over2})e^{-i{\pi\over L}(n+{1\over2})x^{-}}
    +d_{N}^{\dagger}(n+{1\over2})e^{i{\pi\over L}(n+{1\over2})x^{-}}
    \right\}, \\
 Q& = & \sum_{n\ge0}\left[b_{N}^{\dagger}(n+{1\over2})b_{N}(n+{1\over2})
 -d_{N}^{\dagger}(n+{1\over2})d_{N}(n+{1\over2})\right]+N-M(\hat{q}),  \\
 P^{+} & = & {\pi\over L}\sum_{n\ge0}(n+{1\over2})
           \left[b_{N}^{\dagger}(n+{1\over2})b_{N}(n+{1\over2})
                +d_{N}^{\dagger}(n+{1\over2})d_{N}(n+{1\over2})\right] 
                \nonumber \\
       &   &{}+{\pi\over L}\left(-{1\over2}(N-M(\hat{q}))^{2}
              +(N-M(\hat{q}))Q\right).
\end{eqnarray}

We are now ready to present the two-body approximation of the meson 
state,
\begin{equation}
 |K\rangle=\sum_{M=-\infty}^{\infty}
 \int^{M+{1\over2}}_{M-{1\over2}}dq|q\rangle 
 \sum_{k=0}^{K-1}\varphi_{K}(q,k)
 b^{\dagger}_{M}(k+{1\over2})d^{\dagger}_{M}(K-k-{1\over2})|0\rangle_{M},
        \label{meson}
\end{equation}
where $\varphi_{K}(q-1,k)=e^{i\theta}\varphi_{K}(q,k)$.
This state is physical and an 
eigenstate of $P^{+}$ with eigenvalue $(\pi/ L)K$.

In order to obtain the theta dependence of the mass of the meson, we 
need to solve the Einstein-Schr\"odinger equation 
$2P^{+}P^{-}_{\theta}|K\rangle=M^{2}(\theta)|K\rangle$. In the 
two-body sector in which we are working, the Einstein-Schr\"odinger 
equation becomes
\begin{eqnarray}
        & &K\left[
        -{1\over2}{\partial^{2}\over\partial q^{2}}
        +{m^{2}\over2}[\psi({1\over2}+q-{M(q)})+\psi({1\over2}-q+{M(q)})]
        -\tilde\epsilon_{0}(\theta)
        \right]\varphi_{K}(q,k) \nonumber \\
        & &+\left[m^{2}K\left({1\over k+{1\over2}-q+M(q)}+
        {1\over K-k-{1\over2}+q-M(q)}\right)
        +\sum_{l=0,(l\ne k)}^{K-1}{K\over(l-k)^{2}}\right]\varphi_{K}(q,k) 
        \nonumber \\ 
    & &{}+\sum_{l=0}^{K-1}
    \left({1\over K}-{K(1-\delta_{l,k})\over(l-k)^{2}}\right)
    \varphi_{K}(q,l)
    \nonumber \\
    &=&M^{2}(\theta)\varphi_{K}(q,k).
    \label{mesoneq}
\end{eqnarray}
We call this equation the meson equation.

It is instructive to consider the so-called ``continuum'' limit, 
$K\rightarrow \infty$, $L\rightarrow \infty$, keeping the ratio 
$P^{+}=(\pi/ L)K$ finite, of the meson equation. Naively, the
second term of (\ref{mesoneq}) becomes independent of $c\equiv q-M(q)$,
\begin{equation}
        K({1\over n+{1\over2}-c}+{1\over K-n-{1\over2}+c})
        \rightarrow {1\over x}+{1\over 1-x}
\end{equation}
as $K$ goes to infinity with $n/K\rightarrow x$. (Remember $|c| < 
1/2$.) Therefore it might appear that the zero mode decouples
from the non-zero modes,
\begin{eqnarray}
        K\left\{
        -{1\over2}{\partial^{2}\over{\partial q^{2}}}
        +{m^{2}\over 2}[\psi({1\over2}+q-M(q))+\psi({1\over2}-q+M(q))]-
        \tilde\epsilon_{0}\right\}\phi(q,x) & &  \nonumber \\
        +(m^{2}-1)\left({1\over x}+{1\over 1-x}\right)\phi(q,x)
        +\int_{0}^{1} dy (1-{1\over(x-y)^{2}})\phi(q,y) & = & M^{2}\phi(q,x),
        \label{toonaive}
\end{eqnarray}
so that the solution is the product of the solution of the vacuum 
equation $\varphi_{\theta}(q)$ and that of 'tHooft-Bergknoff 
equation\cite{berg77,thooft74},
\begin{equation}
        (m^{2}-1)\left({1\over x}+{1\over 1-x}\right)\Phi(x)
        +\int_{0}^{1} dy (1-{1\over(x-y)^{2}})\Phi(y)  =  M^{2}\Phi(x).
        \label{thooft}
\end{equation}
There is however a subtlety; when $n=0$ and $n=K-1$, the second term 
of (\ref{mesoneq}) is divergent as $c\rightarrow 1/2$ and 
$c\rightarrow -1/2$, respectively. It is therefore not obvious 
whether the zero mode decouples or not. At this moment, we do not know 
if the zero mode really decouples.

\section{Conclusion}
\label{discussion}
In this paper, we treated the zero mode of the gauge field very 
carefully by putting the system in a finite (light-cone) spatial box.
We showed that $\stackrel{o\ }{A_{-}}$ survives Dirac's procedure.
The Hamiltonian in terms of the independent degrees of freedom contains
a complicated interaction term between the fermion and the zero mode.
In order to quantize the model, we carefully defined the current, 
charge, and momentum operators, so that they satisfy the desired 
properties. In particular, we succeeded in constructing the charge 
operator which is invariant under large gauge transformations and 
commutes with the Hamiltonian $P^{-}$. By using it, we were able to 
define the physical space.

A physical state, which annihilates the charge, is a state whose 
fermion Fock state component is related to the zero mode of the gauge 
field. As a very important example, we constructed the vacuum state.
It turned out that it is a linear combination of infinitely many 
$N$-vacua, with the wave function satisfying the vacuum equation 
(\ref{vaceq}). The theta is identified with a Bloch momentum in a 
periodic potential. It is therefore self-evident that the energy 
density is periodic in theta. We proceeded to investigate
the meson state by 
approximating it as a two-body state. We obtained the meson equation 
(\ref{mesoneq}), which determine the theta dependence of the meson 
mass. 

It is interesting to note that the potential of the vacuum equation
(\ref{vaceq}) has singularities at $q$ equal to half-odd integers. 
This singularities stem from the zeros of the Dirac operator 
$D_{-}=\partial_{-}+ie\stackrel{o\ }{A_{-}}$ for anti-periodic 
functions. A proper treatment of 
these zeros might have regularized the singularities of the potential.
Unfortunately, however, we do not understand how to do it.

Numerical solutions of the vacuum equation (\ref{vaceq}) and the 
meson equation (\ref{mesoneq}) are now under study.

\acknowledgments

The authors are very glad to acknowledge the discussions with 
our colleagues in Kyushu University. 
K.~H. would like to thank 
The Ohio-State University, Stanford Linear 
Accelerator Center, and Max-Planck-Institut f\"ur Kernphysik for the 
hospitality, where this work has been worked. He is grateful to 
R.~Perry, S.~Pinsky, D.~G.~Robertson, A.~C.~Kalloniatis, S.~Brodsky, 
M.~V\"ollinger, B.~van~de~Sande, and T.~Heinzl for the discussions. 
M.~T. would like to thank S.~Tanimura and M.~Tachibana for the 
discussions. 
This work 
is supported by a Grand-in-Aid for Scientific Research from the 
Ministry of Education, Science and Culture of Japan (No.06640404).

\appendix

\section{Currents and anomaly}
In this appendix, we discuss the regularization of the current, 
the Schwinger term, and chiral anomaly. In order to have a 
well-defined quantum theory, one must regularize the current properly
so that it reproduces the well-known chiral anomaly.

The massive Schwinger model is a gauge invariant theory. One should 
preserve gauge invariance in any regularization. Actually it is 
possible. On the other hand, axial symmetry is broken anomalously
at the quantum level. There is no consistent way to preserve both
symmetries.

Let us begin with our Fourier expansion of the fermion field 
(\ref{psifourier}). By substituting it into the gauge invariant 
definition of the current (\ref{pointsplitting}), we get
\begin{eqnarray}
{j}^+(x)&=&\sqrt2 :\psi_R^{\dagger}\psi_R(x):_{N} 
     +{1\over2L}(N-q),\\
{j}^-(x)&=&\sqrt2 :\psi_L^{\dagger}\psi_L(x):_{N}
     -{e\over2\pi}\tilde{A^-}, 
\label{jminus}
\end{eqnarray}
in our gauge condition. The normal-ordering is with respect to the 
$N$-vacuum. See eq.~(\ref{normalordered}). 
In deriving these, we used the following 
properties\cite{berg77},
\begin{eqnarray}
        {}_{N}\langle 0| \psi_R^{\dagger}(x+\epsilon)\psi_R(x)|0\rangle_{N} 
        &=& {-i \over 2\sqrt2 \pi}
        {e^{i{\pi\over L}N\epsilon^{-}}\over \epsilon^- -i0}, \\
        {}_{N}\langle 0| \psi_L^{\dagger}(x+\epsilon)\psi_L(x)|0\rangle_{N} 
        &=& {-i \over 2\sqrt2 \pi}
        {e^{i{\pi\over L}N\epsilon^{-}}\over \epsilon^+ -i0}. 
\end{eqnarray}

As explained in the text, we think that the zero mode (the charge) has 
nothing to do with the non-zero modes and ``adjust'' the zero mode 
so that it satisfies desired properties. In Sec. \ref{quantization},
we have constructed such a charge. We only require that 
the nonzero modes of the vector and axial vector currents satisfy the
conservation and anomalous conservation laws respectively.

In order to calculate the divergences of the currents, we need the
commutator of the current, $[\tilde{j}^{+}(x),\tilde{j}^{+}(y)]$.
By a straightforward calculation, we get
\begin{eqnarray}
[\tilde{j}^+(x),\tilde{j}^+(y)]&=&{1\over (2L)^2}
\Big[(\sum_{n=0}^{\infty}\exp\{-i{\pi\over L}(n+{1\over2})(x-y)\})^2  
\nonumber \\
& &{}-(\sum_{n=0}^{\infty}\exp\{i{\pi\over L}(n+{1\over2})(x-y)\})^2 
\Big]
+\cdots, 
\end{eqnarray}
where the ellipsis stands for the operator 
part which vanishes in the ``continuum'' limit. 
Note that the sums do not converge. 
We make them convergent by adding (or 
subtracting) a small imaginary part in the exponents. We get
\begin{eqnarray}
[\tilde{j}^+(x),\tilde{j}^+(y)]&=&{1\over 4\pi}\lim_{\epsilon\rightarrow 0 }
[{1\over (x-y+i\epsilon)^2}-{1\over (x-y-i\epsilon)^2}]+{\cal O}(L^{-2}) 
\nonumber \\
&=& {i\over 2\pi} \delta'(x-y)+{\cal O}(L^{-2}).
\label{schwingerterm}
\end{eqnarray}
In this way, we can reproduce the correct Schwinger term in the 
``continuum'' limit.

It is now easy to calculate the current divergences. By using the 
anomalous commutation relation (\ref{schwingerterm}), one get
\begin{equation}
\partial_+ \tilde{j}^+(x) = -i[\tilde{j}^+(x),P^-]
=im(:\psi_L^{\dagger}\psi_R(x)-\psi_R^{\dagger}\psi_L(x):)_\sim
+{e\over 2\pi}\partial_-\tilde{A^-}.
\end{equation}
{}The spatial derivative of $\tilde{j}^-(x)$ is 
\begin{equation}
\partial_- \tilde{j}^-(x) =
-im(:\psi_L^{\dagger}\psi_R(x)-\psi_R^{\dagger}\psi_L(x):)_\sim
-{e\over 2\pi}\partial_-\tilde{A^-}.
\end{equation}
(This may of course be obtained from the commutator with $P^{+}$ 
defined in Sec.~\ref{mesonsec}.)
{}From these we get the divergences of the vector current and 
the axial vector current: 
\begin{eqnarray}
        \partial_{\mu} \tilde{j}^{\mu}(x)
        &=&\partial_+ \tilde{j}^+(x) + \partial_- \tilde{j}^-(x)
        =0, \\
        \partial_{\mu} \tilde{j_5}^{\mu}(x)
        &=&\partial_+ \tilde{j}^+(x) - \partial_- \tilde{j}^-(x) \nonumber \\
        &=&2im(:\bar{\psi}\gamma_5\psi:)_\sim + {e\over 
        \pi}\epsilon^{\mu\nu}\partial_{\mu}\tilde{A_{\nu}},
        \label{jdiv}
\end{eqnarray}
where we use the relation 
$\gamma^{\mu}\gamma_5=-\epsilon^{\mu\nu}\gamma_{\nu}, (\epsilon^{+-}=-1)$. 

How about the axial charge? As explained in the text, it is formally
equal to the (vector) charge. But because the axial vector current is 
not conserved, we expect that the axial charge is {\it not} conserved.
In conclusion, there is no such a charge on the light-cone. Remember 
that the left-handed field $\psi_{L}$ is not an independent field. 
The independent fields are $\psi_{R}$ and $\stackrel{o\ }{A_{-}}$. It is 
well-known that axial-vector transformations are inconsistent on the 
light-cone\cite{mustaki}, i.e., they are inconsistent with
the constraint equation (\ref{psiconstraint}). 
What if one wants to define the axial-vector
transformations only for the independent field $\psi_{R}$?
Because of $\gamma_{5}\psi_{R}=\psi_{R}$ it is equivalent to the usual
(vector) phase transformations. One cannot define an axial-vector 
transformation, different from the usual
(vector) phase transformation, in a self-consistent way. It means that
the axial charge, which is supposed to be the generator of the 
transformation does not exit.

Mustaki proposed another definition of the axial-vector current which
is conserved even for massive fermions\cite{mustaki}. Does it lead us
to another definition of axial charge? Unfortunately it does not. 
Mustaki's 
conserved current is nothing but the vector current in the massive 
Schwinger model.


\end{document}